# The available force in long-range interaction complex systems and its statistical physical properties

Zhifu Huang[1,*], Congjie Ou[1], Bihong Lin[1], Guozhen Su[2], Jincan Chen[2,*]

[1]College of Information Science and Engineering, Huaqiao University, Xiamen 361021, People's Republic of China

[2]Department of Physics, Xiamen University, Xiamen 361005, People's Republic of China

A new concept of the available force in long-range interaction complex systems is proposed. The relationship between the available force in different time intervals and the interaction parameters of complex systems is described. It is found that when the interaction parameters satisfy a determined condition, the trajectory that the velocity is divergent but the displacement is convergent can be well described and that the long-range interaction, anomalous diffusion, and q-Gaussian type distribution of complex systems can also be well described by the interaction parameters in different cases. In addition, by utilizing the velocity of time series randomly and analyzing its probability distribution of displacement, it is explained that when there exists the long-range interaction in complex systems, the fat-tail distributions will exhibit. The results obtained show that the relationship between the available force and the interaction parameters may be used to investigate the statistical physical properties in long-range interaction complex systems.

________________________________

*Emails: zfhuang@hqu.edu.cn, jcchen@xmu.edu.cn

## 1. Introduction

Characterizing complicated dynamics from experimental time series is a fundamental problem of continuing interest in a wide variety of fields. In experimental systems, the time interval of any process is finite. When the position is ensured, the displacement during the time interval $\Delta t$ can be expressed as

$$x(t, \Delta t) = s(t + \Delta t) - s(t), \qquad (1)$$

where $s(t)$ is the position at time $t$ and $\Delta t$ is the time interval. At the same time, one can define the corresponding velocity as

$$v(t, \Delta t) = \frac{x(t, \Delta t)}{\Delta t} = \frac{s(t + \Delta t) - s(t)}{\Delta t}. \qquad (2)$$

It is important to note that the displacement and velocity mentioned here are not instantaneous values. They depend not only on the time but also on the time interval. Therefore, we can analyze the displacement and velocity in different time intervals.

A typical time interval dependent series is the particle's displacement during anomalous diffusion. Anomalous diffusion is a phenomenon encountered in almost every branch of science. Many physical, biological, and finical systems [1]-[5] that contain fractal and self-similar structures, long-range interaction, and/or long-duration memory, have anomalous diffusions. These anomalous diffusions can be characterized by one-dimensional mean-square displacement as

$$\langle x^2(t, \Delta t) \rangle \propto \Delta t^\alpha, \qquad (3)$$

where $\alpha$ is called the diffusion coefficient. The cases of $\alpha < 1$ and $\alpha > 1$ correspond to the subdiffusion and superdiffusion, respectively. $\alpha = 1$ represents the normal diffusion, which is the result of the Brownian motion. Using equations (2) and (3), one can obtain the mean-square velocity during time interval $\Delta t$ as

$$\langle v^2(t, \Delta t) \rangle \propto \Delta t^{\alpha - 2}. \qquad (4)$$

From equations (3) and (4), it can be seen that when $0 < \alpha < 2$, the mean-square displacement is convergent but the mean-square velocity is divergent at the $\Delta t \to 0$ limit. Since the mean-square velocity is divergent at the $\Delta t \to 0$ limit, the instantaneous velocity can not be strictly measured. Therefore, a moving object may need an infinite force to change its speed or direction discontinuously. In this situation, neither the instantaneous force nor the mean force can be defined. According to Newton's law, we can define an available force as

$$\widetilde{F(t,\Delta t)} = m\frac{v(t+\Delta t,\Delta t)-v(t,\Delta t)}{\Delta t}, \qquad (5)$$

where *m* is the mass of the moving object and we set *m*=1 for the sake of convenience. It is worthwhile to note that in different time intervals, the available force has different values.

## 2. Analysis of an available force

It is well known that different complex systems may exhibit different specific interactions. However, different complex systems sometimes exhibit similar statistical physical properties. In this work, we focus on the change of an available force in different time intervals rather than the specific interaction form of the force in complex systems. The available force in different time intervals can be expressed as

$$\widetilde{F(t_a,\Delta t_a)} = Z[\widetilde{F(t_b',\Delta t_b')}], \qquad (6)$$

where *Z* is the function of the available force in different time intervals. As the specific interaction in a system is complex and sometimes unsolvable, the *Z* function may not be easily obtained. However, an intriguing aspect of complex systems is to exhibit self-similar structures [6] characterized by scale invariance, which plays a central role in a large number of physics phenomena [7]. Therefore, $\widetilde{F(t_a,\Delta t_a)} = \lambda \widetilde{F(t_b',\Delta t_b')}$ can be considered as a specific case of Eq. (6), where $\lambda$ is called the self-similar rate to describe the time-scale invariance property. As mentioned above, the available force depends not only on the time but also on the time interval, so the scale of time and time interval should be identical to keep the self-similarity of the series. A simple case is to assume $k = t_b'/t_a = \Delta t_b'/\Delta t_a = 2$ so

that one can obtain

$$\widehat{F(t,\Delta t)} = \lambda \widehat{F(2t, 2\Delta t)}. \tag{7}$$

The deterministic dynamical systems are usually with a period *T*. Moreover, the time series in complex systems are sometimes nonstationary [8]-[12] so that the single self-similar rate $\lambda$ is insufficient to describe the nonstationary states of complex systems. It may be replaced by multiple self-similar rates $\lambda_i$ $(i=1,2,3...)$. In Eq. (7), $k=2$ is adopted, which indicates that next hierarchy of the time-series is composed of two similar segments. Since the full length of the time-series is confined by the period T, the length of each segment should be *T*/2. In this case, the single $\lambda$ can be replaced by double self-similar rate $(\lambda_1, \lambda_2)$, where $\lambda_1$ and $\lambda_2$ are the self-similar rates of the first segment ($0 \leq t < T/2$) and second segment ($T/2 \leq t < T$), respectively. Therefore, the specific form of the available force in different time intervals can be expressed as

$$\widehat{F(t,\Delta t)} = \begin{cases} \lambda_1 \widehat{F(2t, 2\Delta t)}, & 0 \leq t < \dfrac{T}{2} \\ \lambda_2 \widehat{F(2t-T, 2\Delta t)}, & \dfrac{T}{2} \leq t < T \end{cases}. \tag{8}$$

The values of the self-similar rates $\lambda_1$ and $\lambda_2$ depend closely on the specific interactions among the complex system, and consequently, $\lambda_1$ and $\lambda_2$ can also be called as interaction parameters. For different complex systems, the interactions may be different from each other but it is possible for them to exhibit the same interaction values. This may be helpful to investigate the general properties of complex systems.

It should be mentioned that equation (8) can be iterated by itself to obtain the available force at different time intervals. The corresponding velocity can be derived from the available force and the corresponding time-series of displacement at different time intervals can also be shown. If starting from the iteration at $\Delta t = T/2$ with $\widehat{F(0,T/2)} = \widetilde{F}_0$ and $x(0,T) = 0$ and using $v(0,T) = v(0,T/2) + v(T/2,T/2) = 0$ and equation (5), we can obtain the velocity as $v(0,T/2) = -\widehat{F(0,T/2)}T/4$ and $v(T/2,T/2) = \widehat{F(0,T/2)}T/4$. In addition, the displacements at different times can be expressed as $x(0,T/2) = -\widetilde{F(0,T/2)}T^2/8$ and $x(T/2,T/2) = \widetilde{F(0,T/2)}T^2/8$. By iterating equation (8) with itself, the available forces, velocities and displacements in smaller time intervals ($T/4, T/8, T/16......$) can also be obtained. Without losing generality, we substitute $\widetilde{F_0}=1$ and $T=1$ into equation (8) to simulate the one-dimensional available force of a Brownian-like particle. It is worthwhile to note that $\lambda_1\lambda_2 > 0$ should be avoided during the simulation, otherwise the time-series of equation (8) will always be positive (or negative) to push the particle along one single direction.

Because the available force and velocity are calculated from a large time interval to a small time interval, we can only obtain the discrete velocity series $v(t,T/2^i)$ at $t=0, T/2^i, 2T/2^i, 3T/2^i......$, where $T/2^i$ is the smallest time interval and the integer $i$ represents the iteration times of equation (8). Certainly, when the time interval turns sufficiently small, the velocity can be calculated at any time point. When the corresponding velocity is obtained, the corresponding position at time $t$ and in time interval $T/2^i$ can be written as

$$S(t,\frac{T}{2^i}) = \sum_{j=0}^{\mathrm{Int}(2^i t/T)-1} v(j\frac{T}{2^i},\frac{T}{2^i})\frac{T}{2^i}, \tag{9}$$

where $T/2^i$ is the time step. Equation (9) indicates that the position depend not only on time but also on time interval. The maximum absolute velocity at time interval $T/2^i$ can be derived from equation (8) and given by

$$Max[|v(t,\frac{T}{2^i})|] = \sum_{j=1}^{i}[\frac{Max^{j-1}(|\lambda_1|,|\lambda_2|)\widetilde{F_0}}{2}\frac{T}{2^j}] = \frac{1-[Max(|\lambda_1|,|\lambda_2|)/2]^i}{2-Max(|\lambda_1|,|\lambda_2|)}\frac{\widetilde{F_0}T}{2}. \tag{10}$$

From equation (10), we can find that if $Max(|\lambda_1|,|\lambda_2|) > 2$ and $i \to \infty$, the velocity will diverge. Then, equation (10) can be simplified as

$$Max[|v(t,\frac{T}{2^i})|] = \frac{[Max(|\lambda_1|,|\lambda_2|)/2]^i}{Max(|\lambda_1|,|\lambda_2|)-2}\frac{\widetilde{F_0}T}{2}. \tag{11}$$

Although the velocity diverges, the maximum displacement in time interval $T/2^i$ can be obtained as

$$Max[|x(t,\frac{T}{2^i})|] = Max[|v(t,\frac{T}{2^i})|\frac{T}{2^i}] = \frac{[Max(|\lambda_1|,|\lambda_2|)/4]^i}{Max(|\lambda_1|,|\lambda_2|)-2}\frac{\widetilde{F_0}T^2}{2}. \tag{12}$$

It is found from equation (12) that the displacement is still convergent if $Max(|\lambda_1|,|\lambda_2|) < 4$. Therefore, we can distinguish three different situations by the variation of the interaction parameters. Firstly, if $Max(|\lambda_1|,|\lambda_2|) \leq 2$, the velocity and displacement are both convergent. It may correspond to a smooth trajectory. Secondly, if $2 < Max(|\lambda_1|,|\lambda_2|) < 4$, the velocity is divergent but the displacement is convergent. Finally, if $Max(|\lambda_1|,|\lambda_2|) \geq 4$, the velocity and displacement are both divergent. Our interest is in the second situation that is similar to the real particle motion in various physical systems. Thus, $\lambda_1 = 2.6$ and $\lambda_2 = -1.8$ are adopted during the simulation and the results obtained are shown in Fig.1, where the time interval is decreased from $T/2^{10}$ to $T/2^{12}$, and the position is fixed but the velocity is increased. One can conclude that the velocity of the trajectory is divergent and the displacement of the trajectory is convergent at the $i \to \infty$ limit. It means that the velocity is divergent but integrable everywhere. These behaviors are very similar to the complex systems presenting long-range interaction and/or long-duration memory. Therefore, it is possible to adopt this model to describe different systems in a unified way.

## 3. Statistical properties corresponding to an available force

### 3.1 Long-range correlation of velocity

It is well known that complex systems often include long-range interaction. The velocity autocorrelation function can be written as

$$\rho(n) = \frac{\langle v(t,\Delta t)v(t+n\Delta t,\Delta t)\rangle}{\langle v^2(t,\Delta t)\rangle}, \tag{13}$$

where $n$ is the correlation length. Using equations (5) and (8), we can obtain a series values of velocity. These values can be substituted into Eq. (13) to check the behavior of long-range correlation. As shown in Fig.2, the correlation function of velocity is not equal to zero when the correlation length is large. It is shown explicitly that the long-range interactions are included in the available force. Therefore, the available force is suitable to describe the long-range interaction of complex systems.

**3.2 Anomalous diffusion**

In the case of $Max(|\lambda_1|,|\lambda_2|)<4$, the displacement of equation (12) is convergent at the $i\to\infty$ limit and the exact position at an arbitrary time point exists and can be written as

$$s(t) = S(t,0) = \lim_{i\to\infty}[\sum_{j=0}^{Int(2^i t/T)-1} v(j\frac{T}{2^i},\frac{T}{2^i})\frac{T}{2^i}]. \tag{14}$$

Using equation (14), one can calculate the displacement $x(t,\Delta t)$ for each randomly chosen $t$ and then find the mean square displacement as

$$\sigma^2(\Delta t) = \langle x^2(t,\Delta t)\rangle, \tag{15}$$

As mentioned above, the displacement $x(t,\Delta t)$ varies with $\Delta t$ and $\sigma^2(\Delta t)$ is also a function of the time interval. In order to get a reasonable statistical physical property, $\Delta t$ should be large enough, i.e., $\Delta t >> T/2^i$. Meanwhile, the time interval should be much less than the period, i.e., $T >> \Delta t$. Thus, we adopt $i=27$ and $\Delta t_0 = T/2^{15}$ in the simulation. Using equations (14) and (15), we can plot the curves of $\sigma^2(\Delta t)$ varying with $\Delta t$, as shown in figure 3. The numerical results exhibit excellent agreement with equation (3), where the different choices of $\lambda_1$ or $\lambda_2$ correspond to different values of the diffusion coefficient $\alpha$.

From equation (12), we can see that the velocity is convergent if $Max(|\lambda_1|,|\lambda_2|)<2$. In such a situation, the velocity is convergent and continuous, the trajectory of velocity will be ballistic, and the diffusion coefficient will tend to 2. From equation (12), one can also find that the displacement will diverge at $Max(|\lambda_1|,|\lambda_2|)>4$. Then, the displacements are independent of time interval and the diffusion coefficient will tend to zero. It can be seen from Fig.4 that the diffusion coefficients decrease with the increasing of the interaction parameters $|\lambda_1|$ or $|\lambda_2|$. In fact, Fig. 4 can be filled by more simulation data to construct a transformation table. It means that for a given anomalous diffusion coefficient $\alpha$, one can find a pair of suitable interaction parameters $(\lambda_1, \lambda_2)$, and vice versa. It may be helpful to analyze different kinds of anomalous diffusions and compare their behaviors in a unified way.

**3.3 Probability distribution function of displacement**

The central limit theorem (CLT) is an extremely important concept in probability theory and also lies at the heart of statistical physics [13][14]. It basically says that the sum of $N$ independent identically distributed random variables, rescaled with a factor $1/\sqrt{N}$, has a Gaussian distribution. Suppose that the dynamical system does not satisfy a CLT because it has sufficiently strong correlations, what should the typical probability distributions be for such systems? We can study this question by doing a statistical analysis on the displacements of the motion. Equation (12) makes it possible to calculate the distributions of the displacements for different interaction parameters. The results obtained are shown in figure 5, where $\lambda_1 =$ 2.6 and $\lambda_2 =$ -1.8 and -2.6, respectively. The distribution of displacements are well-fitted by the following $q$-Gaussian function

$$p[x(\Delta t)] \propto [1-(1-q)\beta x^2(\Delta t)]^{1/(1-q)}, \qquad (16)$$

where $\beta$ is a parameter that characterize the width of the distribution and $q$ is the nonextensivity index [15][16]. In Eq. (16), $q \neq 1$ indicates a departure from the Gaussian shape while the $q - 1$ limit yields the normal Gaussian distribution. Figure 5 shows that the values of $q$ depend closely on the interaction parameters $\lambda_1$ and $\lambda_2$. It is shown in figure 6 that different interaction parameters correspond to different values of $q$. It means once again that the relationship between the available force and the interaction parameters may be used to investigate the statistical physical properties in complex systems that have long-range interaction and/or long-duration memory.

Now we can see that the long-range correlation in the time-series of velocity ensures the $q$-exponential distribution of displacement. If we rearrange the velocity series randomly to eliminate the long-range interaction, the distribution of displacement will reduce to the normal distribution. In this case, the displacement can be calculated by

$$x_{ran}(\Delta t) = \lim_{i \to \infty}[\sum_{j=1}^{\text{Int}(\frac{\Delta t}{T/2^i})} v(t_{ran,j}, \frac{T}{2^i})\frac{T}{2^i}], \tag{17}$$

where $v(t_{ran,j}, T/2^i)$ is the discrete value that is randomly chosen from the velocity series. In figure 5, we also plot the distribution of $x_{ran}(\Delta t)$ with $\lambda_1 = 2.6$ and $\lambda_2 = -1.8$. The results obtained can be nicely fitted by the normal Gaussian distribution. It means that normal Gaussian distributions can be generalized to *q*-Gaussian distributions when complex systems exhibit long-range correlations. It is worthwhile to mention that our model coincides with the previous works, such as fractional Brownian motion [17], Lévy motions [18], and fractional stable Lévy motions [19][20], where the fat-tail distributions were encountered when the second moment of velocity is divergent.

The general long-range interaction behavior of complex systems observed in this paper may be of relevance for more general classes of complex systems in nature as well. For example, Caruso et al. [21] observed that the probability distributions of energy differences of subsequent earthquakes in Northern California and in the World Catalog are well fitted by a *q*-Gaussian with *q*=1.75. Their model is just based on long-range interaction, self-organized criticality, and the Olami-Feder-Christensen model. A similar result has also been recently observed for financial markets: A *q*-Gaussian is fit for the interoccurrence times between the losses in stock market for a considerable range of time delays [22]. Moreover, the anomalous diffusion is sometimes associated with *q*-Gaussian distribution, as in the liquid with vortices [23] or in the driven-dissipative dusty plasma [24][25]. We believe that the similarities between our simulation results and the experimental results are not occasional. A general mechanism of long-range interaction deserves a further study by using the available force. This might, in particular, enlighten the deep reasons for the frequent occurrence of long-interaction, anomalous diffusion, and *q*-Gaussians in natural, artificial, financial, and social complex systems.

## 4. Conclusions

To sum up, in order to overcome the divergent of instantaneous velocity, we have defined an available force to describe the long-range interaction in complex systems in a unified way. The trajectory that the velocity is divergent but the displacement is convergent can be well described by the interaction parameters located in $2 < Max(|\lambda_1|, |\lambda_2|) < 4$. Such systems always exhibit long-range interactions, anomalous distributions, and q-exponential distributions. Their statistical properties can be directly simulated by suitable interaction parameters. Moreover, the numerical results support that when there exists the long-range interaction of complex systems, the fat-tail distributions will exhibit. Although further precise calculation is needed to elucidate the connection between the interaction parameters and the specific interactions in complex systems, the present results provide helpful elements for the further understanding of the occurrence of long-range interaction, anomalous diffusion, and q-Gaussian in complex systems. They are also helpful for the correct interpretation of experimental results in some complex dynamical systems including, in particular, the ubiquitous dissipative systems.

**Acknowledgments**

Project supported by the National Natural Science Foundation (No. 11247265, 11005041), the Fujian Provincial Natural Science Foundation (No. 2011J01012, 2010J05007), and the Science Research Fund of Huaqiao University (No.11BS207), People's Republic of China.

**References**

[1] R. N. Mantegna, and H. E. Stanley, Nature **383**, 587 (1996).

[2] R. Metzler, and J. Klafter, Phys. Rep. **339**, 1 (2000).

[3] R. Kutner, A. Pekalski, and K. Sznajd-Weron, *Anomalous Diffusion: From Basics to Applications* (Springer-Verlag, Berlin, 1999).

[4] M. F. Shlesinger, G. M. Zaslavsky, and U. Frisch, *Levy Flights and Related Topics in Physics* (Springer, Berlin, 1995).

[5] J. Szymanski, and M. Weiss, Phys. Rev. Lett. **103**, 038102 (2009).

[6] S. V. Vladimirov, and K. Ostrikov, Phys. Rep. **393,** 175 (2004).

[7] B. B. Mandelbrot, *The Fractal Geometry of Nature* (Freeman, New York, 1982).

[8] J. D. Hamilton, Econometrica, **57**, 357 (1989).

[9] C. K. Peng, S. Havlin, H. E. Stanley, and A. L. Goldberger, Chaos **5**, 82 (1995).

[10] C. Govern, and P. R. ten Wolde, Phys. Rev. Lett. **109**, 218103 (2012).

[11] R. Chelakkot, R. G. Winkler, and G. Gompper, Phys. Rev. Lett. **109**, 178101 (2012).

[12] A. Dechant, and E. Lutz, Phys. Rev. Lett. **108**, 230601 (2012).

[13] N. G. van Kampen, *Stochastic Processes in Physics and Chemistry* (North-Holland, Amsterdam, 1981).

[14] A. Y. Khinchin, *Mathematical Foundations of Statistical Mechanics* (Dover, New York, 1949).

[15] C. Tsallis, J. Stat. Phys. **52**,479 (1988).

[16] C. Tsallis, *Introduction to Nonextensive Statistical Mechanics: Approaching a Complex World* (Springer, New York, 2009).

[17] B. B. Mandelbrot, and J. W. van Ness, SIAM（Society for Industrial and Applied Mathematics）Rev. **10,** 422 (1968).

[18] P. Lévy, Théorie de l'addition des Variables Aléatoires (Gauthier-Villars, Paris, 1954).

[19] M. S. Taqqu, and R. Wolpert, Z. Wahrscheinlichkeitstheorie verw. Gebiete **62**, 53 (1983).

[20] M. Maejima, Z. Wahrscheinlichkeitstheorie verw. Gebiete **62**, 235 (1983).

[21] F. Caruso, A. Pluchino, and V. Latora, Sergio Vinciguerra and Andrea Rapisarda, Phys. Rev. E **75**, 055101(R) (2007).

[22] J. Ludescher, C. Tsallis and A. Bunde, Euro. Phys. Lett., **95**, 68002 (2011).

[23] S. Ratynskaia, K. Rypdal, C. Knapek, S. Khrapak, A. V. Milovanov, A. Ivlev, J. J. Rasmussen, and G. E. Morfill, Phys. Rev. Lett. **96**, 105010 (2006).

[24] B. Liu, and J. Goree, Phys. Rev. Lett. **100**, 055003 (2008).

[25] B. Liu, and J. Goree, Phys. Rev. E **78**, 046403 (2008).

[26] B. Liu, and J. Goree, Phys. Rev. E **75**, 016405 (2007).

Figure captions:

Fig.1. The illustration of velocity (a, b) and displacement (c, d) versus time for the parameters $\lambda_1 = 2.6$ and $\lambda_2 = -1.8$, where the cases of (a, c) and (b, d) correspond to the cases of $\Delta t = T/2^{10}$ and $T/2^{12}$, respectively.

Fig.2. The autocorrelation function of velocity versus *n* curves for $\lambda_1 = 2.6$ and $\lambda_2 = -1.8$.

Fig.3. The mean square displacement versus time interval curves for $\lambda_1 = 1.5$.

Fig.4. The diffusion coefficient versus $-\lambda_2$ curves.

Fig.5. The distributions of dimensionless displacements. The standard Gaussian curve and *q*-Gaussian curves with *q* = 1.59 and 0.83 are represented by dashed and solid lines, respectively.

Fig.6. The *q* versus $-\lambda_2$ curves.

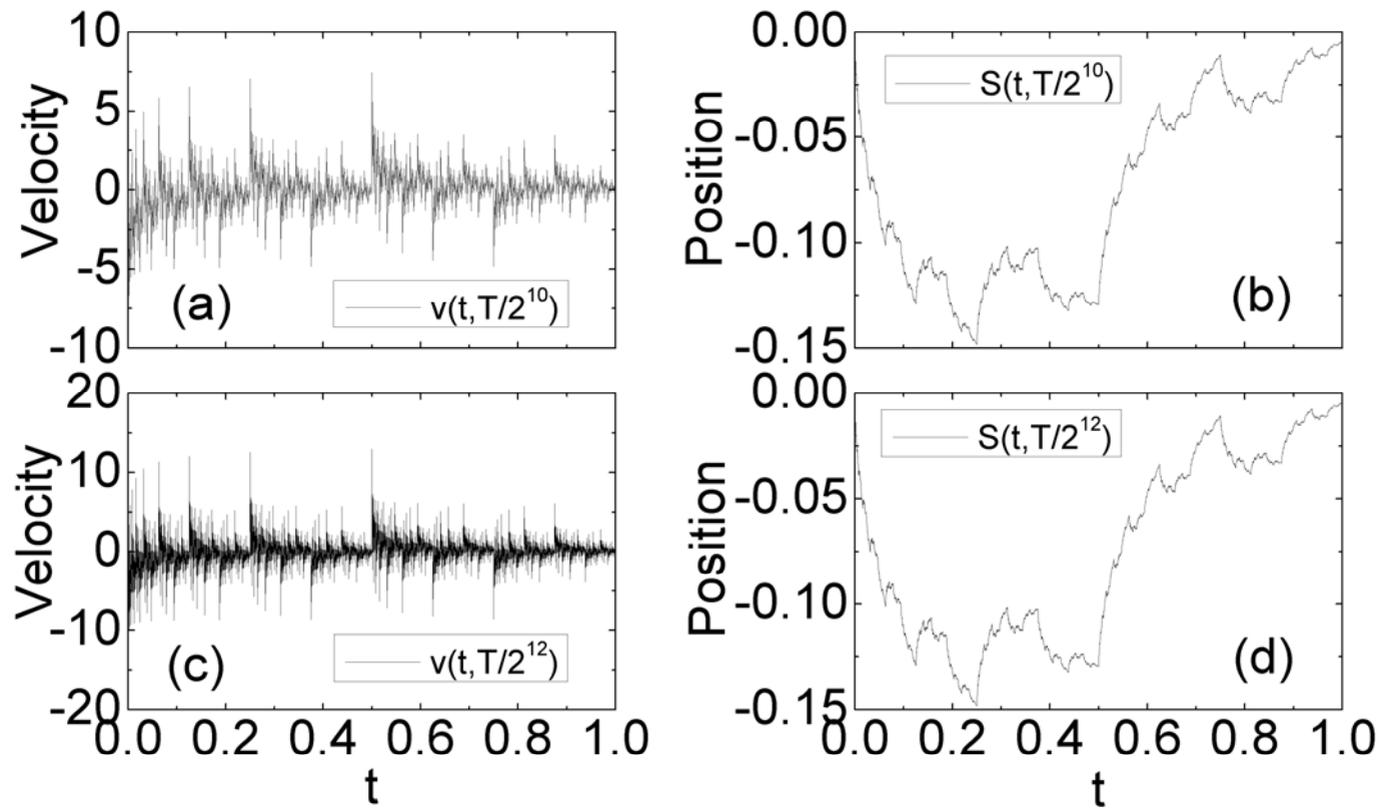

Fig.1.

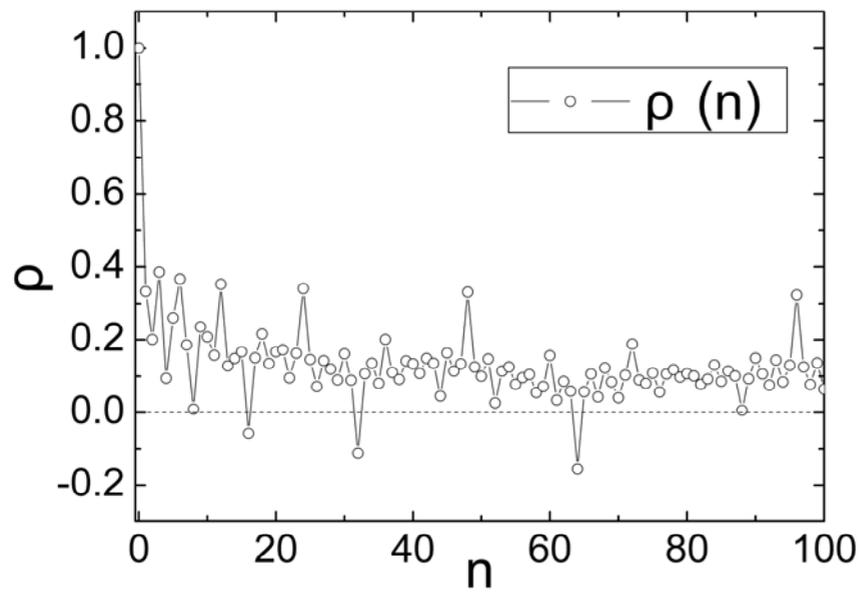

Fig.2.

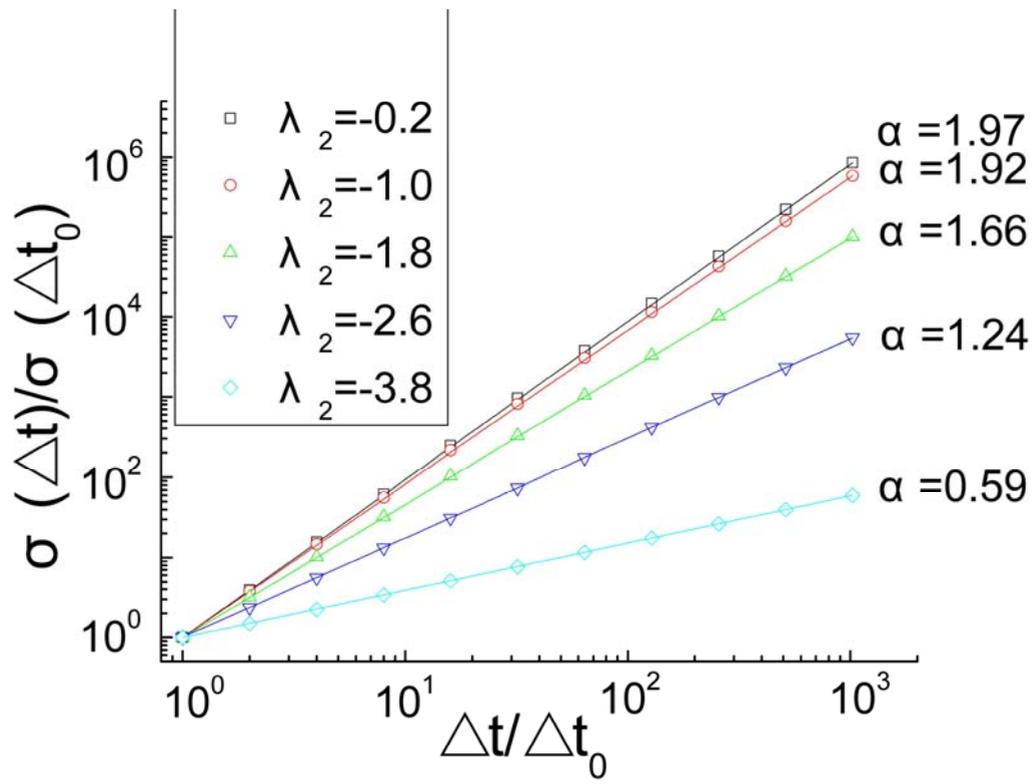

Fig.3.

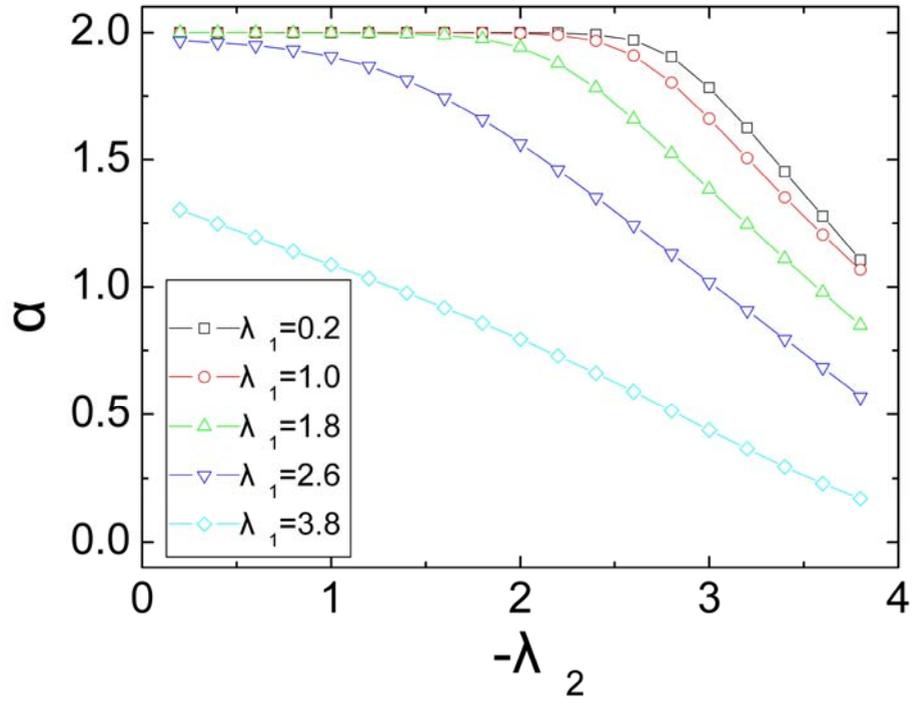

Fig.4.

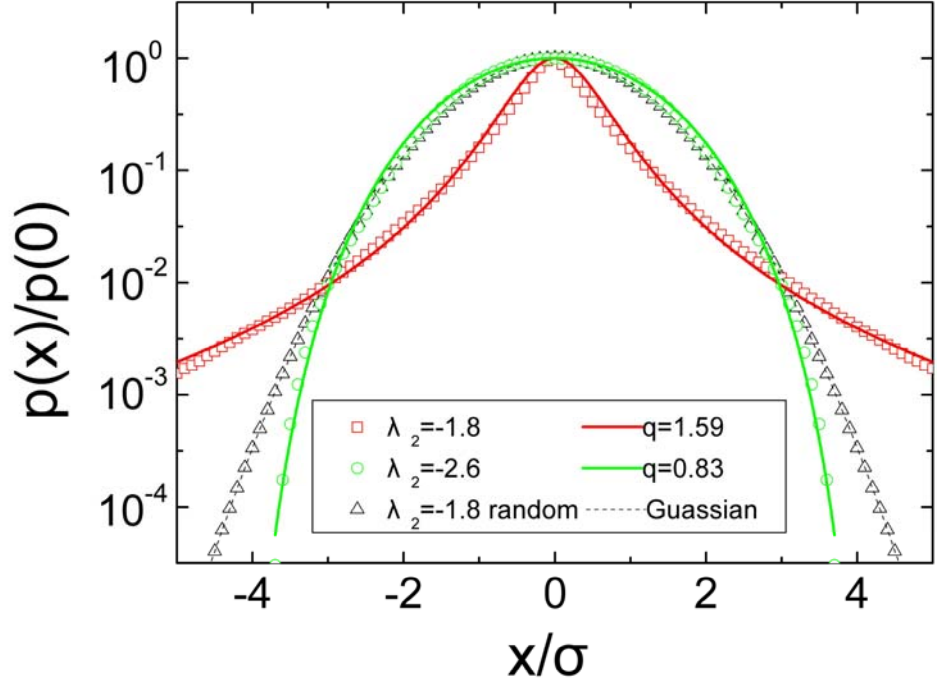

Fig.5.

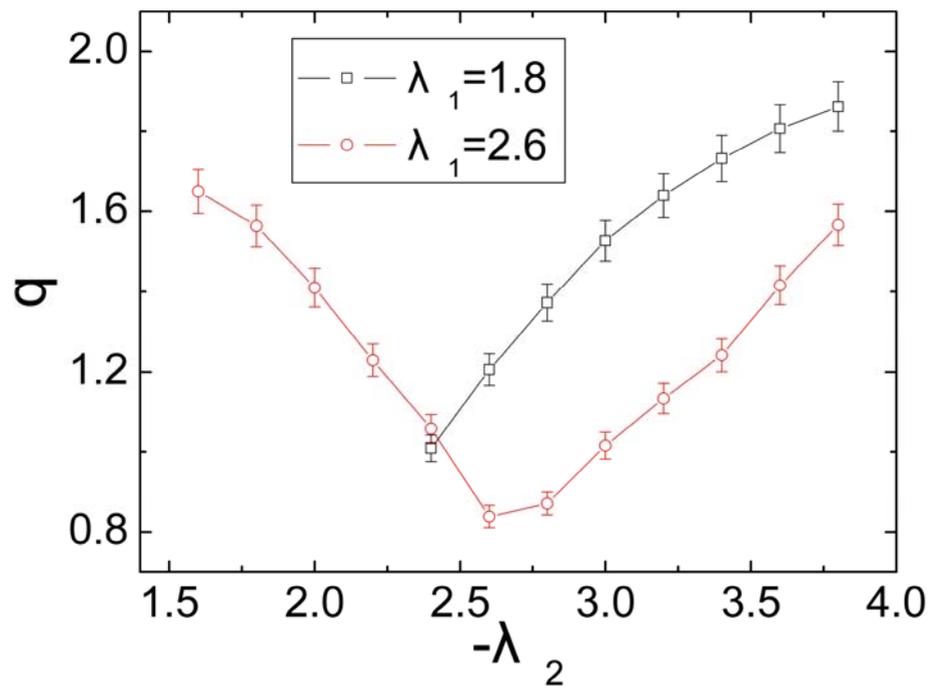

Fig.6.